\documentstyle[12pt,colordvi,epsf]{article}

\begin{document}
\renewcommand{\thefootnote}{\fnsymbol{footnote}}
\begin{center}
{\LARGE BARYON NUMBER PENETRABILITY AS A MEASURE OF ISOTHERMAL BARYON NUMBER 
FLUCTUATIONS IN THE EARLY UNIVERSE}\vspace{.3in}\\
{\large B. K. Patra$^{1,}$\footnote[2]{E-mail: bkpatra@iopb.stpbh.soft.net} 
, V. J. Menon$^2$ and C. P. Singh$^2$ }\vspace{.2in}\\
{\it {$^1$}Institute of Physics, Sachivalaya Marg, Bhubaneswar 751 005, India\\
$^2$Department of Physics, Banaras Hindu University, 
Varanasi 221 005, India.}\vspace{.2in}\\
\underline{Abstract}
\end{center}

We have examined the efficiency of baryon-number transport mechanism across 
the phase boundary
in a cosmological quark-hadron phase transition through the proper estimate of
baryon-number penetrability $\langle \Sigma_h \rangle $. For this purpose we 
have derived first the double-pair creation probability $P_b$ in terms 
of single-pair creation probabilty per unit time and unit volume $\kappa_m$ 
and then obtained an analytical expression for $\langle \Sigma_h \rangle$. Our calculation is free from the
uncertainty of the value of double-pair creation probability per unit
time and unit volume $\kappa_b$ which was used as a free
parameter in earlier calculations. Finally the variations of double-pair
creation probability $P_b$ as well as $\langle
\Sigma_h \rangle $ with temperature are shown and compared with other known 
results. 

\pagebreak
\section{\bf Introduction} 
Recently there is a growing interest in the primordial nucleosynthesis in an 
Universe having an inhomogeneous distribution of baryon number. The most 
popular mechanism for the generation of baryon number inhomogeneities has been 
suggested as a first-order QCD phase transition [1-8]. The present Universe is 
presumed to have undergone several successive phase transitions associated with symmetry breaking in its early stage. A QCD phase transition might have 
occurred at approximately 10$^{-6}$ sec after the big bang in the 
early Universe. Even though many aspects of phase transition are not a 
settled issue, yet, the QCD phase transition is expected to be a 
first-order one from lattice QCD results [9]. Therefore, a large local 
fluctuation in the baryon to photon ratio $n_B/n_\gamma$ could arise which 
might subsequently modify the standard picture of the primordial 
nucleosynthesis (PNS). Such a fluctuation might also result into the formation 
of quark nuggets and/or quark stars etc.[1-8]. The basis for the production of 
isothermal baryon number
fluctuation lies in the separation of cosmic phases as follows [1,5]. Initially
the Universe is in the quark-gluon plasma (QGP) phase at a high temperature and
the net baryon number resides entirely in the quark-gluon plasma phase and is 
distributed homogeneously. As the Universe expands, the temperature drops to 
the critical temperature $T_c$ where the quark-gluon plasma exists in thermal 
and chemical equilibrium with the dense and hot hadron gas. Subsequently, the 
expansion requires a continuous conversion of QGP into the hadron phase. The 
phase transition is completed when all the quark-gluon plasma has been 
converted to hadron phase and all the baryon number residing finally in the 
hadron phase is distributed homogeneously. The magnitude of baryon number 
fluctuation is estimated by the baryon contrast ratio $R_{eq}$ of the net 
equilibrium baryon number density in the quark-gluon plasma (QGP) phase to 
that in the hadron gas (HG) phase, i.e., $R_{eq} =n^B_{QGP}/n^B_{HG}$ which is 
evaluated at the critical temperature and chemical potential. The baryon number density inhomogeneity arising due to such a quark-hadron phase transition will 
thus alter the yields of the PNS which is a function of $R_{eq}$ and the 
neutron to proton $(n/p)$ ratio [10]. Several theoretical attempts have been 
recently made to determine the values of $R_{eq}$ [11-14]. 
\par All aforementioned calculations of $R_{eq}$ assume that the Universe is in 
complete thermal and chemical equilibrium during the phase transition [15]. But in 
reality, the cosmic first-order phase transition will necessarily result in 
deviations from thermal and chemical equilibrium because the low temperature 
hadron phase is not nucleated immediately at the critical temperature T$_c$. A 
generic feature of quantum or thermal nucleation theory is that the nucleation 
rate does not become large unless the temperature has dropped below $T_c$ [16]. 
The magnitude of these deviations will depend on the efficiency of heat and 
baryon transport across the transition front during the phase transition. 
Latent heat transport is carried out by the motion of the boundary
wall which converts volume of one vacuum into another. This vacuum energy 
difference acts as a source for particle creation. However, the latent heat or 
entropy could also be carried out across the phase boundary by neutrinos apart 
from surface evaporation of hadrons (mostly pions). Baryon number is not
thermally created in the hadron phase. Therefore, it actually flows across the
boundary by the strong interaction physics if the hadron phase is going to
have any net baryon number at all.
\par There are two limiting situations governing the efficiency of 
baryon number transport across the phase boundary [5.17]. In the first situation, when
the bubble-like regions of quark-gluon plasma are small in size, the
phase boundaries move slowly compared to the baryon number transport rate. 
Hence the baryon number is quickly and efficiently transported across the
boundaries, so as to maintain the chemical equilibrium. In the second 
situation, the boundaries of the bubble-like regions move rapidly compared to 
the baryon transport rate so that the net baryon transport process becomes 
inefficient. Therefore, chemical equilibrium may not be achieved on the time 
scale of the phase coexistence evolution and, consequently, the baryons could 
be 
concentrated in the shrinking bubble-like regions of the QGP. The magnitude of 
the resulting baryon number fluctuation at the time of phase decoupling will 
still be given by the ratio $R^\prime$ (say) of baryon number densities in the
two phases but now each phase will have a different baryon chemical
potentials. Thus, the final ratio $R^\prime$ will be larger than the 
equilibrium ratio $R_{eq}$.
\par It is clear that in both the abovementioned situations isothermal
baryon number fluctuations will result. However, the magnitude of $R_{eq}$ is 
in the limit of efficient baryon number transport mechanism which is the 
aforesaid first situation. The efficiency of baryon number transport across 
the phase interface will be given by the bulk properties of the phases and the 
estimate of baryon number penetrability [5,17]. The baryon number penetrability 
$\Sigma_h$ from the hadronic side is defined as the probability that a baryon 
which approaches from the hadronic phase will dissociate into quarks and pass 
over into the QGP phase. Similarly from another side, we could define the 
probability $\Sigma_q$ that the quarks which approaches towards the phase 
boundary will form a color singlet baryon and pass over to the hadron phase.
 The thermal average of these probabilities are related by the detailed balance
\begin{eqnarray}
k~f_{q-\overline q} \langle \Sigma_q \rangle = f_{b-\overline b}
\langle \Sigma_h \rangle
\end{eqnarray} 
where $f_{q-\overline q}~(f_{b-\overline b})$ is the excess quark flux (baryon
flux) in the QGP (HG) phase, respectively, and the dimensionless quantity $k$ 
takes values from 1/3 to 1. The value $k = 1/3 $ implies that baryon number
predominantly passes over the phase boundary as preformed three-quark
clusters in the QGP phase. Similarly $k =1 $ signifies that baryons will
be predominantly formed by the leading quarks which cross over the phase 
boundary into the hadron phase. A low value for $\langle \Sigma_h \rangle$ 
(or $\langle \Sigma_q \rangle $) ($< 1$) implies an early drop out of chemical 
equilibrium during the separation of phases and thus might lead to the
large  amplitude of baryon-number fluctuations. 
\par The estimation of baryon number penetrability has been attempted 
recently [5,17,18] by many authors within the framework of the
chromoelectric flux tube (CFT) model. Here an energetic leading quark in the
QGP is assumed to pick up an antiquark forming thereby an expanding CFT
whose decay through a single or double pair creation results in the formation
of mesons or baryons, respectively. Naturally, the calculation of
$\langle \Sigma_h \rangle$ within CFT model requires the values of the
single as well as coherent double pair creation probabilities per unit
space per unit time, called $\kappa_m$ and $\kappa_b$, respectively. 
Although $\kappa_b$ is a crucial factor for the formation of a baryon
yet no theoretical expression for it based on strong interaction physics
exists in the literature.\\
\par In this context the following points pertaining to a recent paper
by Jedamzik and Fuller [17, referred to as JF hereafter] become
particularly relevant : (i) JF have extracted the numerical value of
$\kappa_b$ by analyzing empirically the ratio of baryons to mesons
produced in the $e^+ e^- \rightarrow $ hadrons experiment conducted
in the accelerator laboratories. (ii) JF allow for the full angular
range $0 \le \theta \le \pi$ where $\theta$ is the polar angle
of the leading quark's velocity with respect to the normal to the
phase boundary. (iii) The final expression for $\langle \Sigma_h \rangle $
written as an integral over the quark distribution functions is computed
by JF entirely numerically. (iv) Finally, the threshold energy in
JF's work increases with the temperature in an unreasonable manner.\\ 
\par The aim of the present paper is to extend/modify the theory
of JF [17] in multifold respects : (i) By regarding the unconnected double-pair 
creation event as a succession of two single-pair events, we show in Sec.2
below, that the double-pair creation probability $P_b$ (over finite time
duration) can be obtained in terms of the $\kappa_m$ parameter itself.
(ii) Since the leading quark's velocity should be directed towards
the phase boundary, we point out in Sec.3 that the corresponding
polar angle should be restricted to the range $0\le \theta \le \pi/2$.
(iii) We derive in Sec.3 an analytical expression for the thermal
average of the baryon number penetrability so that the dependence
of $\langle \Sigma_h \rangle $ on the relevant variables becomes more
transparent. (iv) We show in Sec.4 the comparison between our
and JF models by plotting $P_b$ as well as $\langle \Sigma_h\rangle$
as functions of the temperature when the quarks are assigned either
the constituent or current mass. (v) Finally, we carefully analyze
the temperature-dependence of the threshold energy in Sec.4 to obtain 
an elegant formula for $\langle \Sigma_h \rangle$ in the high T limit.

\section{\bf Decay Statistics of Flux Tubes}
Chromoelectric flux tube (CFT) models provide us a phenomenological description
regarding the formation of a hadron from quarks and gluons [19]. These models assume
the existence of a chromoelectric field between two oppositely coloured
quarks where the chromoelectric field strength is assumed to be constant
in magnitude and independent of the separation of the quarks. These fields
can be thought of as being confined to a tube of constant width known as
flux tube. Lattice QCD results justify the existence of such a flux tube [21].
Initially, chromoelectric flux tube models were used to explain the
processes such as $e^+e^- \rightarrow $ hadrons [22,23]. Later these models
have also been used to describe the spectrum of mesonic and baryonic
resonances  [22,24].  Recently flux tube models have been employed to estimate
the meson evaporation rates from quark-gluon plasma produced in relativistic
heavy-ion collisions [20] and also to estimate baryon-number penetrability
across the phase boundary in cosmological quark-hadron phase transition [17,18].
\par If the CFT is regarded as an unstable system then its
stochastic properties can be conveniently discussed by introducing the 
following abbreviations : SP $\equiv$ survival probability, DP $\equiv$ decay
probability, m $\equiv$ mesonic or single-pair production channel, b $\equiv$
baryonic or connected double-pair production channel, $b^\ast
\equiv $ disconnected double-pair production channel. Let us now take up the
the earlier model used by Jedamzik and Fuller [17] and a new proposal due 
to us.
\vspace{.2in}\\
\subsection{\bf Jedamzik and Fuller (JF) Model}
In analogy with QED, at a perturbative field theoretic level, the JF model 
amounts to connected diagrams shown in Figs. 1(m) and 1(b). The corresponding
DP's per 
\begin{figure}[h]
\begin{center}
\leavevmode
\epsfysize=7truecm \vbox{\epsfbox{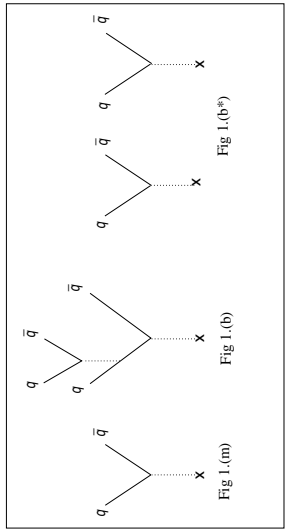}}
\end{center}
\caption{(m), (b), ($b^\ast$) : Diagrams drawn in analogy with 
perturbative QED, representing decay into single, connected double, and
unconnected double $q \overline q$ pair channels, respectively. The crosses
stand for the chromoelectric field as a source. The dotted lines denote
gluons.}
\label{}
\end{figure}
unit time per unit volume are denoted by $\kappa_m$ and $\kappa_b$, 
respectively, having a total $\kappa = \kappa_m + \kappa_b$. The space-time
integrated parameters are called as
\begin{eqnarray}
w_m & = &  \kappa_m \int_{0}^{t_o}  V(t) dt  \\
w_b & = & \kappa_b \int_{0}^{t_o}  V(t) dt 
\end{eqnarray}
and $w  =  w_m + w_b $, $V(t)$ is the instantaneous volume of the flux 
tube and $t_0 \sim E/\sigma$ is the maximum time upto which it expands 
when the incident quark has energy $E$. From the 
conservation of energy and momentum parallel to the phase boundary, the above 
integral could be written as in [17]
\begin{eqnarray}
\int_{0}^{t_o}  V(t) dt = \frac{\pi \Lambda^2}{2\sigma^2}~E^2 \cos\theta
\end{eqnarray}
with $\theta$ being the angle of incidence of the leading quark relative to the 
normal to the phase boundary, $\Lambda$, the radius of the flux tube and  
$\sigma$ the string tension of the flux tube.\\ 
Then, according to the classical theroy of radioactivity, the net SP and DP 
counting all channels would be 
\begin{eqnarray}
P_s & = &  e^{-w} \\
P_d & = &  1 - e^{-w} 
\end{eqnarray}
Since the relative weights for the $m$ and $b$ channels are $w_m/w$ and 
$w_b/w$, respectively, the corresponding channelwise DP's are as used by JF : 
\begin{eqnarray}
P_m & = & \frac{\kappa_m}{\kappa}~P_d  \nonumber\\
    & = & \frac{\kappa_m}{\kappa}~\left( 1-exp \left[-\kappa
\int_{0}^{t_o}  V(t) dt \right] \right) \\
P_b & = & \frac{\kappa_b}{\kappa}~P_d  \nonumber\\
    & = & \frac{\kappa_b}{\kappa}~\left( 1-exp \left[-\kappa
\int_{0}^{t_o}  V(t) dt \right] \right) 
\end{eqnarray}
Although $\kappa_m$ can be estimated from Schwinger's single-pair production
mechanism [25] applied to QCD, nothing a priori is known about $\kappa_b$.
However, JF have extracted empirically the magnitude of $\kappa_b$ from the 
analysis of $e^+e^- \rightarrow$ hadrons data.
\subsection{\bf Our Proposal}
Suppose we retain information about the single $q \overline q$ pair production
of Fig. 1(m) but ignore the diagram 1(b). Then, the double-pair production
event may be looked upon as a sequence of two unconnected single-pair creations
within time $t_0$ as  shown in Fig.1($b^\ast$). Since the flux tube is now 
characterized only by the parameter $w_m$, hence radioactivity theory would 
give the following 
expressions for the net SP and DP over the duration time $t_0$ :
\begin{eqnarray}
P_s^\ast & = & e^{-w_m} \\ 
P_d^\ast & = & 1 - e^{-w_m} = \sum_{n=1}^\infty p_n^\ast 
\end{eqnarray}
where the DP of having exactly $n$ successive single-pair events is a Poissonian
viz.
\begin{eqnarray}
p_n^\ast = e^{-w_m} \frac{w_m^n}{n!}
\end{eqnarray}
Therefore, in our approach the DP for the single and double- pair would be
obtained from
\begin{eqnarray}
P_m^\ast = p_1^\ast = e^{-w_m}~w_m
\end{eqnarray}
\begin{eqnarray}
P_b^\ast = p_2^\ast = e^{-w_m}~\frac{w_m^2}{2}
\end{eqnarray}
In contrast to the JF approach $(7, 8)$, our proposal involves only one 
parameter $w_m$ (or $\kappa_m$) as an input.

\section{\bf Analytical Estimation of Baryon number Penetrability}

The main result of JF's work is contained in their baryon penetrability
equation which gives the thermal average of the 
baryon penetrability integrated over the $q \overline q$ distribution function. We wish to make two plausible comments on this result. Firstly, the angular
integration range in their calculation $0 \le \theta \le \pi$ may not be 
appropriate because if the leading quark is to enter the phase boundary then 
it should have $ 0 \le \theta \le \pi/2$. Therefore we suggest that the 
corrected expression should be 
\begin{eqnarray}
\langle \Sigma_h \rangle = \frac{1}{f_{b-\overline b}} \int_{0}^{\pi/2} d\theta
\int_{E_{th}}^{\infty} dE ~\frac{dn_{q-\overline q}}{dE d\theta}~ {\dot{x}_\perp}^q
(\theta) ~P_b(E,\theta)
\end{eqnarray}
Here $f_{b-\overline b}$, the excess baryon flux in the hadron phase, is given 
by [17]
\begin{eqnarray}
f_{b-\overline b}~ = ~ \frac{\mu_b}{T}~\frac{g_b}{2\pi^2}~(m_b+T)~T^2~exp(-m_b/T)
\end{eqnarray}
where $m_b$, $\mu_b$ and $g_b$ are the baryon mass, baryon chemical potential 
and degeneracy factors, respectively and ${\dot{x}_\perp}^q$ is the 
component of the leading quark velocity perpendicular to the phase boundary,
and $E_{th}$ is the threshold energy.\\
Secondly, JF have done their subsequent calculations numerically so that the
dependence of $\langle \Sigma_h \rangle $ on the relevant parameters
remains somewhat obscure. We suggest that an analytical estimate for the 
seemingly complicated expression (14) is very desirable because that would make
the dependence of $\langle \Sigma_h \rangle $ on the relevant parameters more
transparent. For this purpose we proceed as follows :\\
Using Boltzmann approximation(E/T $\gg$ 1), the differential excess quark 
number density in a given energy interval $dE$ and in a given 
interval of incident angles $d \theta$ becomes
\begin{eqnarray}
\frac{dn_{q-\overline q}}{dE d\theta}~ \sim ~ \frac{\mu_q}{T}~\frac{g_q}
{2\pi^2}~E^2~exp(-E/T)~\sin\theta
\end{eqnarray}
where $\mu_q$, $g_q$ are the quark chemical potential, and the statistical
weight of quarks, respectively.\\
Then Eq.(14) becomes
\begin{eqnarray}
\langle \Sigma_h \rangle \approx C ~\int_{0}^{1}ds ~s~\int_{E_{th}}^{\infty}
dE ~E^2~exp(-E/T)~ P_b(E,\theta)
\end{eqnarray}
where
\begin{eqnarray}
C~\equiv~ \frac{1}{f_{b-\overline b}}~ \frac{\mu_q}{T}~ \frac{g_q}{2\pi^2}
~;~ s = \cos \theta
\end{eqnarray}

Due to the rapidly damped factor $exp(-E/T)$ in Eq.(17) most contribution to 
the energy integration comes from around the threshold energy $E_{th}$. 
Therefore, it is convenient to make the transformation
\begin{eqnarray}
\rho = \frac{E-E_{th}}{T}~;~ E =  E_{th} \{ 1 + \frac{T}{E_{th}} \rho \}
\end{eqnarray}
Upon using the identity $ \int_{0}^{\infty}d\rho~exp(-\rho) = 1 $, Eq.(17) 
reduces to
\begin{eqnarray}
\langle \Sigma_h \rangle  \approx 
 ~C ~\int_{0}^{1}ds ~s~T~E_{th}^2~exp(-\frac{E_{th}}{T})~P_b(E_{th},\theta)
\end{eqnarray}
within correction terms of order $T/E_{th}$. Now, at large angles $\theta 
\rightarrow \pi/2$, $s \rightarrow 0$, $E_{th} \rightarrow \infty$ implying
that $exp(-E_{th}/T)$ gets heavily damped. Hence, most contribution to the
angular integral must come from around the forward direction $\theta 
\rightarrow 0$, $s \rightarrow 1$. For convenience define
\begin{eqnarray}
\hat{E_{th}} &\equiv& {\left. E_{th}\right|}_{\theta =0} \nonumber\\
&=& m_b + B n_q^{-1}
\end{eqnarray}
\begin{eqnarray}
\hat{P_b} & \equiv& P_b(\hat{E_{th}}, \theta=0)
\end{eqnarray}
\begin{eqnarray}
\lambda \equiv \frac{E_{th}-\hat{E_{th}}}{T}  \nonumber\\
~\approx~\frac{m_b}{T} (1-s)
\end{eqnarray}
Here $B$ is the bag constant and $n_q$ is the quark plus antiquark density.
The above eq.(21) is the threshold energy condition in the forward angle
direction $(\theta =0)$ and the next eq.(22) is the corresponding probability.
Using the identity $ \int_{0}^{m_b/T}d\lambda~exp(-\lambda)\approx 1 $, eq.(20) 
yields 
\begin{eqnarray}
\langle \Sigma_h \rangle & \approx & ~\frac{C~T^2~{\hat{E_{th}}^2}}{m_b}
~exp(-~\hat{E_{th}}/T) ~\hat{P_b}~ \{ 1+O(\frac{T}{m_b})\} 
\end{eqnarray}
Substituting Eq.(18) for $C$, we arrive at the desired analytical estimate :
\begin{eqnarray}
{\langle \Sigma_h \rangle}~ =~\frac{\mu_q}{\mu_b}~ \frac{g_q}{g_b} ~
{\left( \frac{\hat{E_{th}}}{m_b} \right)}^2~ e^{-(\hat{E_{th}} -m_b)/T}~ \hat{P_b}
\{1+O(\frac{T}{m_b})\}
\end{eqnarray}
The probability function in JF model
\begin{eqnarray}
\hat{P_b}~&=&~a~\left[1~-~exp \{-b~ \hat{E_{th}}^2 \} \right]
\end{eqnarray}
can be replaced in our model as
\begin{eqnarray}
\hat{P_b^\ast}~&=&~e^{-w_m}~\frac{w_m^2}{2}
\end{eqnarray}
with
\begin{eqnarray}
a~&=&~\frac{\kappa_b}{\kappa_m + \kappa_b}~;~b~=~(\kappa_m +\kappa_b)~
\frac{\pi\Lambda^2}{2 \sigma^2}\\
w_m~&=&~\kappa_m~\frac{\pi\Lambda^2 \hat{E_{th}^2}}{2\sigma^2}
\end{eqnarray}
Here the string tension of the flux tube $\sigma \simeq 0.177
GeV^2$, and single-pair creation probability per unit time per unit
volume $\kappa_m$ is given by [25]
\begin{eqnarray}
\kappa_m = \frac{\sigma^2}{4\pi^3}~exp(-\pi m_q^2/\sigma)
\end{eqnarray}
where $m_q$ is the quark mass. Since there is an ambiguity in choosing the 
precise value of $m_q$, we allow in the sequel both the possibilities
viz. $m_q=300 MeV$ (constituent quark mass) and $m_q=10 MeV$ (current
quark mass). Eqs.(25), (26), (27) form the main algebraic results of the 
present paper.

\section{\bf Results and Discussions}
In Fig. 2, we have shown the variation of double pair creation probability 
 at threshold in the forward direction with the temperature for a small value 
of $b=0.047~GeV^{-2}$ corresponding to the constituent quark mass 
$(m_q=330 MeV)$. Here we have taken the value of the ratio $a (\approx \kappa_m/
\kappa_b) \approx 1/5$ as in [17] 

\begin{figure}[h]
\begin{center}
\leavevmode
\epsfysize=10truecm \vbox{\epsfbox{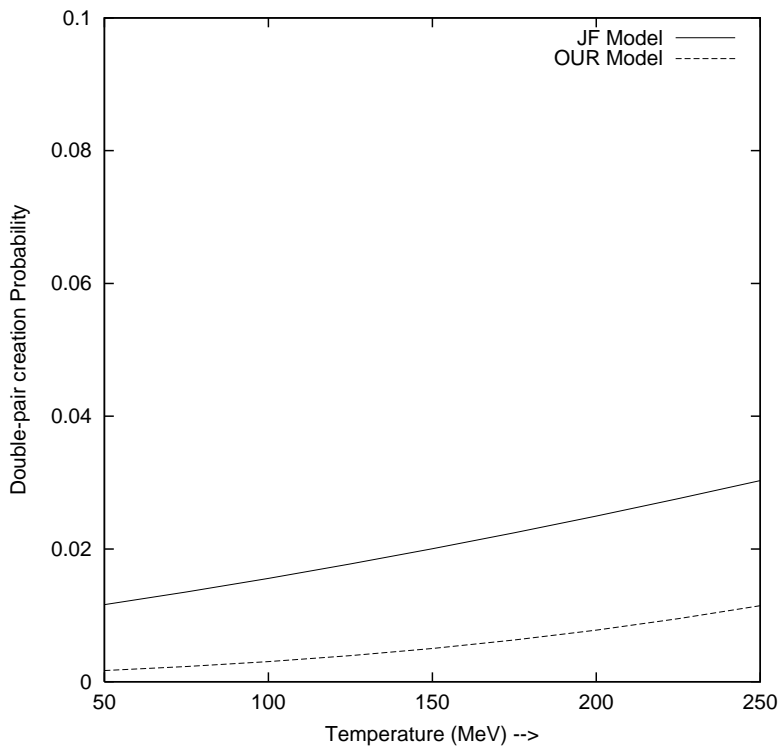}}
\end{center}
\caption{ Variation of double pair creation probability $\hat{P_b}$ 
($\hat{P_b^\ast}$) at threshold in the forward direction 
with the temperature T when the quark has constituent mass. The dashed curve
represents our model calculation (cf. Eqs.(13), (27)) whereas the solid
line represents the JF model (cf. Eqs.(8), (26)). For remaining symbols see
Eqs.(28,29,30)}
\label{}
\end{figure}

Fig.3 is also the same as Fig.2 except that a higher value of $b=0.32~GeV^{-2}$ is taken
corresponding to the current quark mass $(m_q=10 MeV)$ and the ratio $a 
\approx 1/20$.

\begin{figure}[h]
\begin{center}
\leavevmode
\epsfysize=10truecm \vbox{\epsfbox{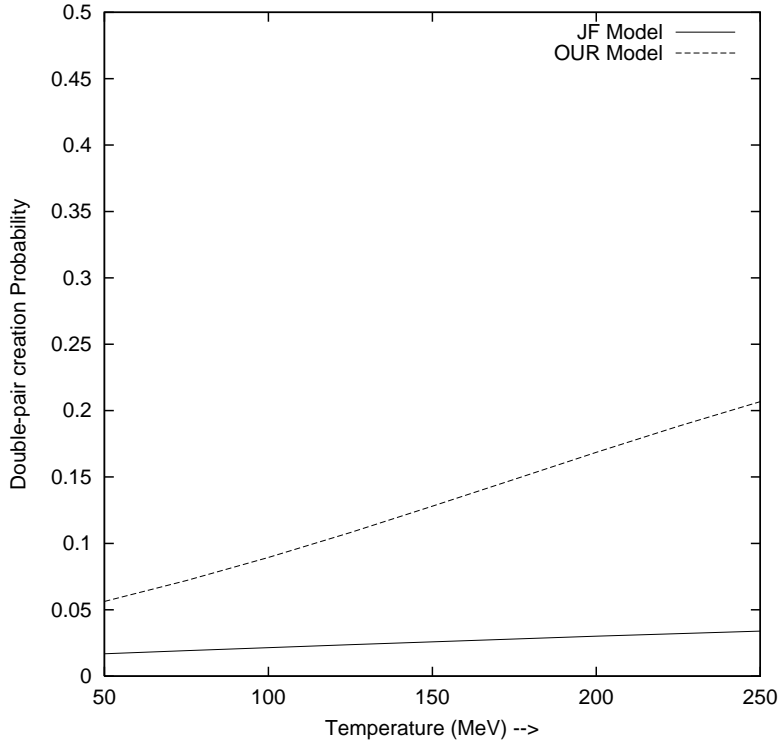}}
\end{center}
\caption{ Same as Fig.2 except that current quark mass has been chosen in
both models.}
\label{}
\end{figure}

Clearly, the difference between the predicted values of $\hat{P_b}$ in our
model and JF model must be attributed to their different premises. Our
approach has the advantage of working with only one decay parameter 
$\kappa_m$ but ignores the connected diagram 1(b). The JF approach has the
advantage of including the graph 1(b) but at the cost of bringing-in
the additional unknown decay parameter $\kappa_b$.
\par Next, we turn to the calculation of the baryon number penetrability
$\langle \Sigma_h \rangle$ based on the analytical result (25). The
probability $\hat{P_b}$ can have either the JF form (26) or our form
(27) with the remaining parameters having been set as for u, d flavours
in QGP sector and nucleons in the hadronic sector. Figs. 4 and 5
display the dependence of $\langle \Sigma_h \rangle$ on the temperature
when the quark has the constituent and current masses, respectively.

\begin{figure}[h]
\begin{center}
\leavevmode
\epsfysize=10truecm \vbox{\epsfbox{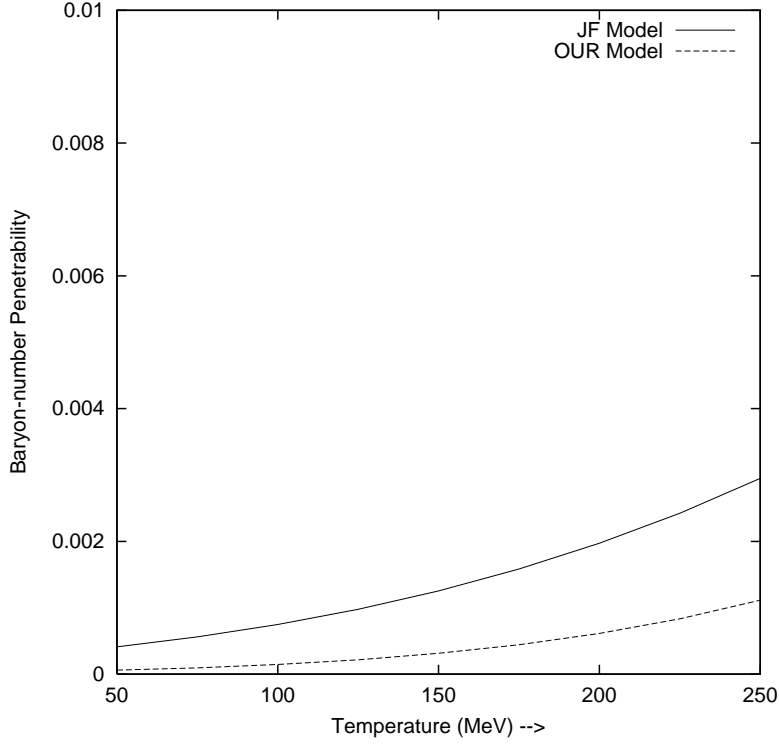}}
\end{center}
\caption{ Variation of baryon number penetrability $\langle \Sigma_h \rangle$ 
(cf. Eqs.(25), (26), (27)) with the temperature T when the constituent quark
mass is chosen. Other notations are the same as in Fig.2}
\label{}
\end{figure}

\begin{figure}[h]
\begin{center}
\leavevmode
\epsfysize=10truecm \vbox{\epsfbox{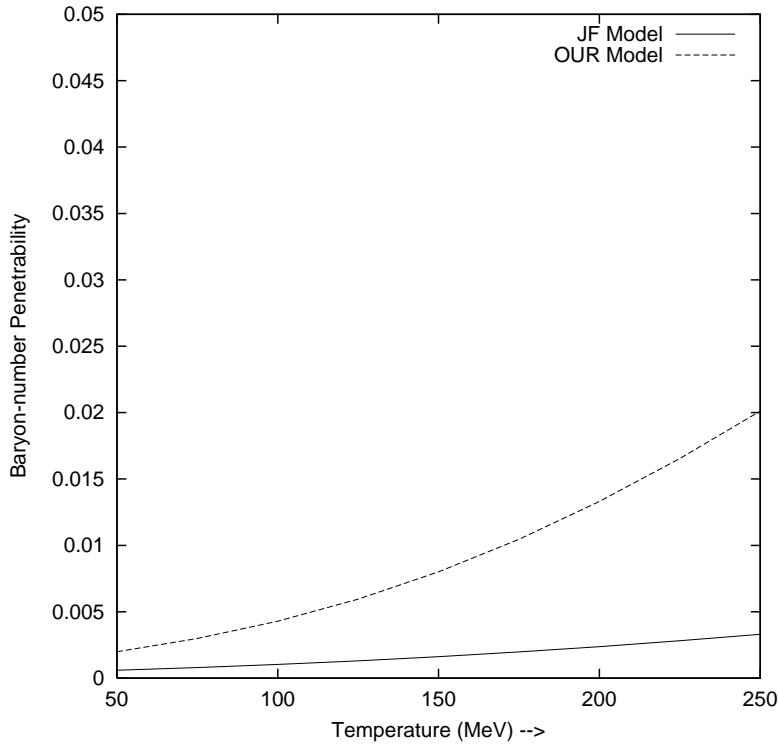}}
\end{center}
\caption{ Same as fig.4 except that the current quark mass is chosen.}
\label{}
\end{figure}

\par In Fig.4, when the quark has constituent mass, our curve lies below that
of the JF model. Our small predicted value $\langle \Sigma_h \rangle \sim
O(10^{-3})$, incidentally, justifies the results obtained by Sumiyoshi
et al.[18] and Fuller et al.[5]. However, in the opposite limit, i.e., for 
current quark mass, Fig.5 reveals that our graph is consistently above that of 
JF. Our predicted higher value of $\langle \Sigma_h \rangle \simeq O(10^{-2})$ 
seems to agree with the original result obtained by Jedamzik and Fuller [17] 
using numerical quadrature involving the unphysical angular range $0 \le 
\theta \le \pi$.
\par In passing we note that the typical time upto which the CFT expands
is $t_0 \sim \hat{E_{th}}/\sigma \sim 10^{-24} s $ which seems to be
one order of magnitude smaller than typical nuclear time scale of $10^{-23} s$. Can the $b^\ast$ channel of Fig.1 be important under such a circumstance? To
answer this question we borrow the $w_m$ values from Figs. 2-5 and look at the
ratio
\begin{eqnarray}
\frac{P_b^\ast}{P_m^\ast} = \frac{w_m}{2}~\sim~2\% ~ {\rm to}~15 \%
\end{eqnarray}
which is sizable. Therefore, our view of regarding the double-pair production
event as a succession of two single-pair events can be justified.
\par Towards the end we wish to comment on a couple of factors on which
baryon number penetrability crucially depends within the chromoelectric
flux tube approach. One factor is the choice of the quark mass as seen
from Figs. 4 and 5. Another factor, in which $\langle \Sigma_h \rangle$
is very sensitive, is the threshold energy $\hat{E_{th}}$ in the
forward direction. Following the idea of Jedamzik and Fuller [17], this threshold 
energy $E_{th}$ consists
of two parts : one is simply the rest mass of the baryon $m_b$ and the other
is the interaction energy $Bn_q^{-1}$ for each quark and antiquark which
resides in the QGP. JF have parametrized the interaction energy as $\approx
3.7 T$ which apparently grows with the temperature. However, in our opinion, 
the function $Bn_q^{-1}$ should decrease with $T$ like $T^{-3}$  for a 
fixed value of the bag constant $B$ since the total quark density 
$n_q$ is an increasing function of $T$. So $E_{th}$ would tend to
$m_b$ in Eq.(25) which yields the elegant high-temperature behaviour
\begin{eqnarray}
\langle \Sigma_h \rangle \stackrel{T \rightarrow \infty}{\longrightarrow} \hat{P_b}
\end{eqnarray}
remembering that $g_q =12$, $g_b=4$ and $\mu_b=3\mu_q$.

\noindent {\bf ACKNOWLEDGEMENTS} : 
One of us, VJM thanks the UGC, Goverment of India, New Delhi for financial support.

\pagebreak

\end{document}